\documentclass[reprint, amsmath, amssymb, aps]{revtex4-1}

\usepackage[pdftex]{graphicx}
\usepackage{wasysym}
\usepackage{amsfonts}
\usepackage{amsthm}
\usepackage{ifsym}
\usepackage{ulem}

\newcommand{\COMMENT}[1]{}

\newcommand{\neqa}{\nonumber\end{eqnarray}}

\newcommand{\ur}[1]{(\ref{#1})}

\def\tr{{\rm tr~}}

\newcommand{\<}{{\langle}}
\renewcommand{\>}{{\rangle}}

\newcommand{\rH}{\mathrm{H}}

\newcommand{\re}{\relax{\rm I\kern-.18em R}}

\def\su2{{SU(2)}}

\def\[{\left[}
\def\]{\right]}

\def\({\left(}
\def\){\right)}
\def\[{\left[}
\def\]{\right]}

\def\<{\langle}
\def\>{\rangle}

\def\i2{\frac{i}{2}}

\def\2F1{\,_2{\rm F}_1}

\newtheorem*{theorem}{Theorem}
\newtheorem*{lemma}{Lemma}

\usepackage[usenames,dvipsnames]{xcolor}
\usepackage{color}
\usepackage{graphicx}
\usepackage{dcolumn}
\usepackage{bm}
\usepackage{hyperref}

\newcommand{\beq}{\begin{equation}}
\newcommand{\eeq}{\end{equation}}
\newcommand{\beqq}{\begin{equation*}}
\newcommand{\eeqq}{\end{equation*}}
\newcommand\beqa{\begin{eqnarray}}
\newcommand\eeqa{\end{eqnarray}}
\newcommand\beqaa{\begin{eqnarray*}}
\newcommand\eeqaa{\end{eqnarray*}}
\newcommand\bea{\begin{array}}
\newcommand\eea{\end{array}}

\begin{document}

\title{Diagonal Form Factors and Hexagon Form Factors}

\author{Yunfeng Jiang$^{\displaystyle\mathcal{F}}$ and Andrei Petrovskii$^{\displaystyle\rho}$}

\affiliation{
\vspace{5mm}
$^{\,\displaystyle\mathcal{F}}$Institute for Theoretical Physics, ETH Z\"urich, 8093 Z\"urich, Switzerland\\
$^{\,\displaystyle\rho}$Institut de Physique Th\'eorique, CEA Saclay, F91191 Gif-sur-Yvette, France
}

\begin{abstract}
\vspace{1mm}
We study the heavy-heavy-light (HHL) three-point functions in the planar $\mathcal{N}=4$ super-Yang-Mills theory using the recently proposed hexagon bootstrap program \cite{Basso:2015zoa}. We prove the conjecture of Bajnok, Janik and Wereszczynski \cite{Bajnok:2014sza} on the polynomial $L$-dependence of HHL
structure constant up to the leading finite-size corrections, where $L$ is the length of the heavy operators. The proof is presented for a specific set-up but the method can be applied to more general situations.
\end{abstract}

\maketitle

\section{Introduction}
Structure constants or OPE coefficients constitute an essential part of data which characterize a conformal field theory. In the case of planar $\mathcal{N}=4$ Super-Yang-Mills theory (SYM), it proves to be fruitful to make connection between the structure constant and the form factors in 2d integrable field theories. This can be most easily done for the heavy-heavy-light (HHL) three point function \cite{Zarembo:2010ab,Costa:3pt} where the two heavy operators are regarded as `incoming' and `outgoing' states and the light operator as an operator sandwiched between these states. Based on this intuition, Bajnok, Janik and Wereszczynski (BJW) \cite{Bajnok:2014sza} conjectured that the $L$-dependence ($L$ is the length of the heavy operators) of HHL structure constant takes the same form as the volume dependence of the diagonal form factors in 2d integrable field theories \cite{Pozsgay:2007gx} \textit{at any coupling}. This conjecture was confirmed by some examples at strong coupling \cite{Bajnok:2014sza} and proved at the leading and one-loop order in the $\mathfrak{su}(2)$ sector at weak coupling \cite{Hollo:2015cda}.\par

Very recently, a non-perturbative method for computing structure constants in $\mathcal{N}=4$ was proposed by Basso, Komatsu and Vieira (BKV)\cite{Basso:2015zoa} called the hexagon bootstrap program. This approach offers us a powerful tool to investigate the HHL structure constant at finite coupling and test further the BJW conjecture. In this paper, we initiate a systematic study of HHL structure constant using the hexagon approach and prove the BJW conjecture at higher loop orders. The proof is presented for a specific set-up for simplicity, but our method can be applied to more general cases \cite{HJP}.

\section{The hexagon bootstrap program}
We review briefly the proposal by Basso, Komatsu and Vieira \cite{Basso:2015zoa}. Intuitively, the structure constant can be represented by a pair of pants either in the spin chain or string theory description. The main idea of \cite{Basso:2015zoa} is cutting this pair of pants into two objects called the hexagons or the hexagon form factors. When cutting the pair of pants, the excitations on each operator can be attributed to either hexagon and one needs to sum over all the possible partitions, each partition is associated with certain weight. In addition, gluing back the two hexagons into a pair of pants requires summing over all possible states living on the three gluing segments, which can be performed by means of integrating over the mirror excitations. The calculation of the hexagons can be achieved in two steps. The first step is to move all excitations on the same edge by performing mirror transformations. The resulting object is called fundamental hexagon. It can be denoted as $\rH^{A_1\dot{A}_1\cdots A_N\dot{A}_N}(u_1,\cdots,u_N)$, where $A_i\dot{A}_i$ are the $\mathfrak{su}_L(2|2)\otimes\mathfrak{su}_R(2|2)$ bifundamental indices for the $i$-th excitation and $u_i$ is the corresponding rapidity. The second step is the computation of the fundamental hexagon itself which is given by the following prescription
\begin{align}
\rH=\rH^{\text{dyn}}\cdot\rH^{\text{mat}},
\end{align}
where
\begin{align}
&\rH^{\text{dyn}}=\prod_{i<j} h(u_i,u_j)\\\nonumber
&\rH^{\text{mat}}=(-1)^{\mathfrak{f}}\,\langle\chi_N^{\dot{A}_N}\cdots \chi_1^{\dot{A}_1}|\mathcal{S}|\chi_1^{A_1}\cdots\chi_N^{A_N}\rangle,
\end{align}
and the factor $(-1)^\mathfrak{f}$ accommodates for the grading. For our case below, we consider only scalar excitations and $\mathfrak{f}=0$.
The dynamical part is simply a product of the scalar function $h(u,v)$ given by
\begin{align}
h_{12}=\frac{x_1^--x_2^-}{x_1^--x_2^+}\frac{1-1/x_1^-x_2^+}{1-1/x_1^+x_2^+}\frac{1}{\sigma_{12}}
\end{align}
where the variables $x_{1,2}^\pm$ are defined as $x_1^\pm=x(u\pm i/2)$ and $x_2^\pm=x(v\pm i/2)$. Here $x(u)$ is the Zhukowsky variable satisfying $x+1/x=u/g$ and $\sigma_{12}=\sigma(u,v)$ is the square root of BES dressing phase\cite{Beisert:2006ez}.\par

For the matrix part, $\chi^A$ denotes a state in the fundamental representation of $\mathfrak{su}(2|2)$ and $\mathcal{S}$ is Beisert's $S$-matrix  \cite{Beisert:2005tm,Beisert:2006qh}. The matrix part of the hexagon form factor is thus given by a factorized product of Beisert's $S$-matrix elements with the dressing phase set to 1.

\section{Set-up}
We study the HHL structure constant with the excitations of the two heavy operators being the transverse scalar excitations $X=\Phi_{1\dot{1}}$ and $\bar{X}=-\Phi_{2\dot{2}}$ and the light operator being the BPS operator, which is also called the reservoir in \cite{Basso:2015zoa}:
\begin{align}
\mathcal{O}_3=\tr\tilde{Z}^{2l_0},\qquad \tilde{Z}=Z+\bar{Z}+Y-\bar{Y}.
\end{align}
The two heavy operators are made of the scalar fields $\mathcal{O}_1:\{Z,X\}$ and $\mathcal{O}_2:\{\bar{Z},\bar{X}\}$. These are the operators in the so-called $\mathfrak{su}(2)$ sector. The length of the heavy operators are $L_1=L_2=L$.
In the heavy-heavy-light three-point function, we have $l_0\ll L$. In this paper, we consider the asymptotic $L$-dependence of the HHL structure constant. This means we neglect the wrapping corrections to the states (physical wrapping), which start to contribute at order $\mathcal{O}(g^{2L})$. However, there are another kind of wrapping corrections, corrections to the correlator itself, referred to as bridge wrapping in \cite{Basso:2015zoa}. On account of the small length of the light operator these corrections contribute at very low orders and should be taken into account.
\begin{figure}[h!]
\begin{center}
\includegraphics[scale=0.4]{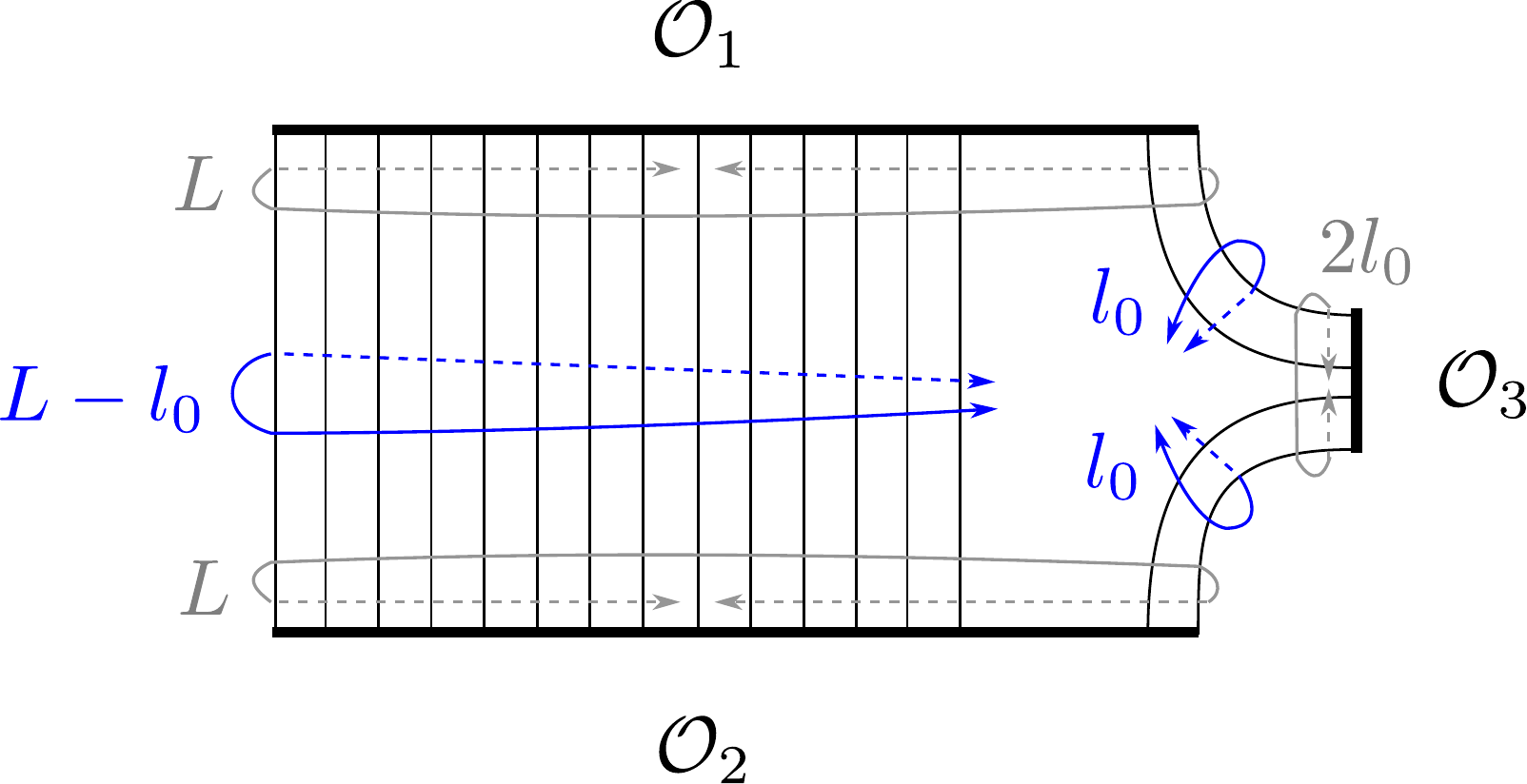}
\caption{Two kinds of wrapping corrections of the structure constant. The blue lines corresponds to the bridge wrappings and the gray lines corresponds to the state wrappings.}
\label{fig:wrapping}
\end{center}
\end{figure}
We postpone this question to future investigation \cite{HJP}. In this paper, we assume $1\ll l_0\ll L$ so that we can trust the result up to relatively high orders without worrying about the wrapping corrections. In this regime, we only need to consider the physical excitations and we can study the asymptotic $L$-dependence. We will see the result confirms the BJW conjecture.\par

\begin{figure}[h!]
\begin{center}
\includegraphics[scale=0.4]{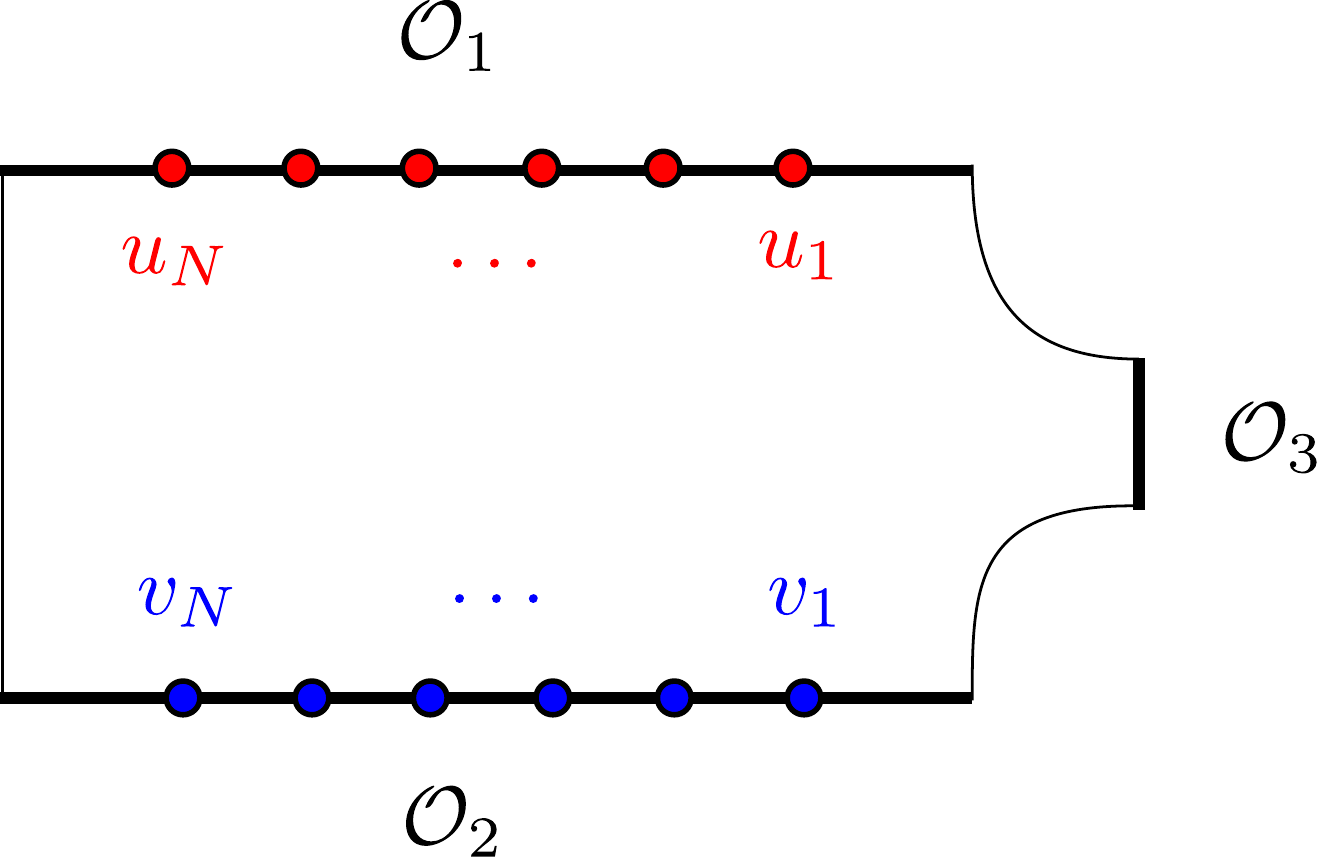}
\caption{The rapidities for the two heavy operators. Notice that the for $\mathcal{O}_1$, the set rapidities is labeled by $u_N,u_{N-1},\cdots,u_1$ while for $\mathcal{O}_2$ is labeled by $v_1,v_2,\cdots,v_N$.}
\label{fig:rap}
\end{center}
\end{figure}

Let us denote the two sets of rapidities of the excitations on $\mathcal{O}_1$ and $\mathcal{O}_2$ to be $\{u\}_N=\{u_N,\cdots,u_1\}$ and $\{v\}_N=\{v_1,\cdots,v_N\}$, as is shown in FIG.\,\ref{fig:rap}.
The structure constant has the following sum-over-partition expression \cite{Basso:2015zoa}
\begin{align}
\label{eq:sop}
\mathcal{C}_{2N}=\sum_{\alpha\cup\bar{\alpha}=\{u\}_N\atop \beta\cup\bar{\beta}=\{v\}_N} \omega_{-l}(\alpha,\bar{\alpha})\omega_l(\beta,\bar{\beta})\rH(\alpha|\beta)\rH(\bar{\beta}|\bar{\alpha}).
\end{align}
where the two splitting factors are
\begin{align}
\label{eq:sfactor}
\omega_{-l}(\alpha,\bar{\alpha})=&\,\prod_{u_j\in\bar{\alpha}}(e^{-ilp(u_j)}\prod_{u_i\in\alpha\atop i>j} S(u_i,u_j))\\\nonumber
\omega_l(\beta,\bar{\beta})=&\,\prod_{v_j\in\bar{\beta}}(e^{ilp(v_j)}\prod_{v_i\in\beta\atop i>j} S(v_j,v_i)).
\end{align}
and $l=L-l_0\sim L$. Here $\rH(\alpha|\beta)$ and $\rH(\bar{\beta}|\bar{\alpha})$ are the hexagon form factors which can be computed non-perturbatively. Note that we have applied Bethe Ansatz Equations (BAE) to rewrite the splitting factor in (\ref{eq:sfactor}). Then both splitting factors $\omega_{-l}$ and $\omega_l$ depend on the large size scale $l$. This is the origin of the explicit $L$-dependence of the structure constant. As we will see later, when we take the diagonal limit $\{v\}_N\to \{u\}_N$, a $\frac{0}{0}$ uncertainty appears, so we will have to take derivatives of the phase factors $e^{ilp(v)}$, which leads to the polynomial dependence of $L$. Another source of the $L$-dependence is the phase factors itself, but, after taking the limit, it can be eliminated by applying BAE.

\section{The hexagon form factor}
\begin{figure}[h!]
\begin{center}
\includegraphics[scale=0.25]{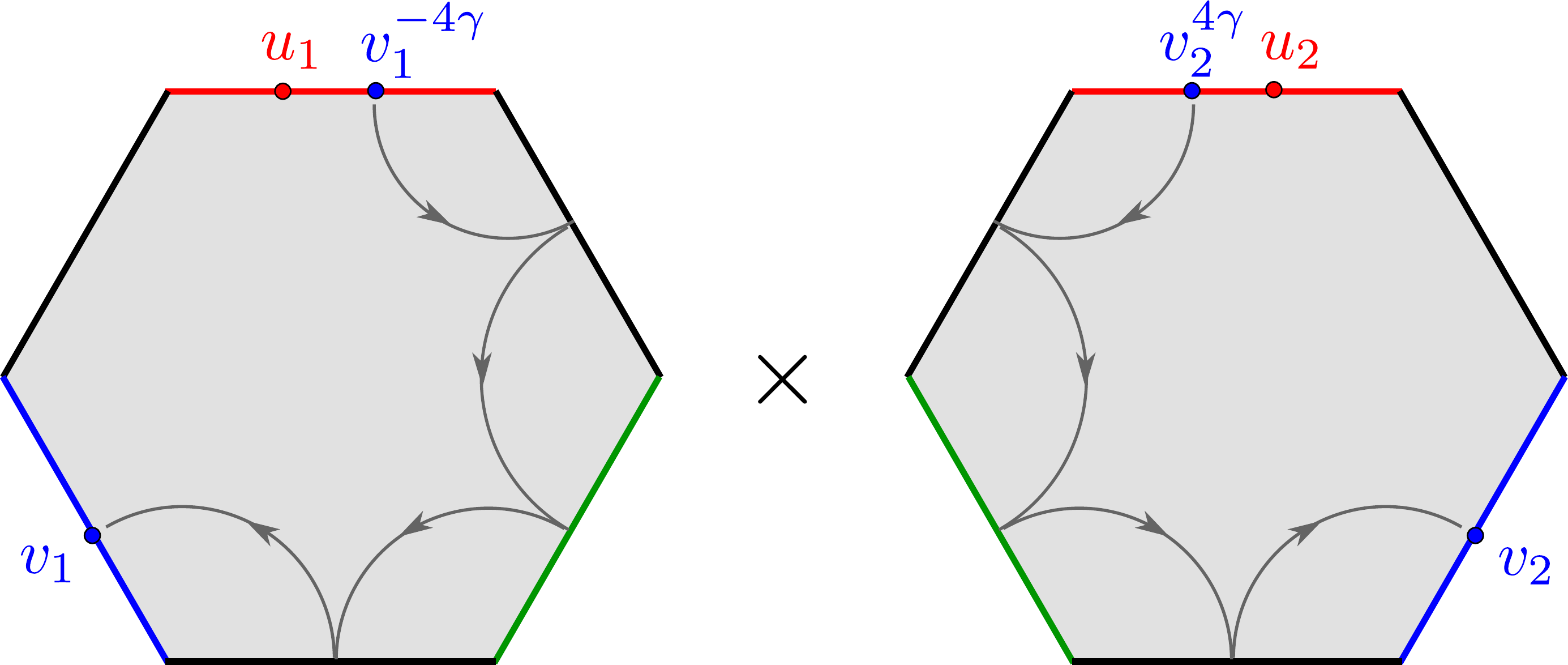}
\caption{Crossing transformation for the excitations of the two hexagons.}
\label{fig:2magnon}
\end{center}
\end{figure}

We first analyze the structure of hexagon form factor and show that there are kinematic poles in the diagonal limit. Each term in the sum-over-partition formula (\ref{eq:sfactor}) contains the product of two hexagon form factors. In order to perform the computation, one needs to perform crossing transformations to move all the excitations on the same edge. Here we choose a $-4\gamma$ transformation for the first hexagon and a $4\gamma$ transformation for the second one, as is shown in FIG.\,\ref{fig:2magnon}.
The advantage of performing $\pm 4\gamma$ transformations is that they leave the matrix part of the hexagon invariant and the kinematic pole in the diagonal limit only appears in the dynamical part. Since the dynamical part takes a factorized form, it is much easier to keep track of the kinematic poles. When performing the crossing transformation in the spin chain frame, in general an extra phase factor appears. In our case, if we perform the $\pm 4\gamma$ transformation, this phase factor is trivial. The dynamical parts of the two hexagons are given by
\begin{align}
&\rH^{\text{dyn}}(\{u\}_n|\{v^{-4\gamma}\}_n)\\\nonumber
&=\prod_{k=1}^n\frac{u_k-v_k+i}{u_k-v_k}\frac{1}{\tilde{h}(v_k,u_k)}\frac{\prod_{i<j}h(u_i,u_j)h(v_i,v_j)}{\prod_{j\ne k}h(v_i,u_j)},\\\nonumber
&\rH^{\text{dyn}}(\{v^{4\gamma}\}_n|\{u\}_n)\\\nonumber
&=\prod_{k=1}^n\frac{u_k-v_k-i}{u_k-v_k}\frac{1}{\tilde{h}(u_k,v_k)}\frac{\prod_{i<j}h(u_i,u_j)h(v_i,v_j)}{\prod_{j\ne k}h(u_i,v_j)}.
\end{align}
where we have used the fact that $h(u,v^{-4\gamma})=1/h(v,u)$ and $h(v^{4\gamma},u)=1/h(u,v)$ and
\begin{align}
\frac{1}{h(u,v)}=\frac{u-v-i}{u-v}\frac{1}{\tilde{h}(u,v)}.
\end{align}
Here
\begin{align}
\tilde{h}(u,v)=\frac{(1-1/x_1^-x_2^+)^2}{(1-1/x_1^-x_2^-)(1-1/x_1^+x_2^+)}\frac{1}{\sigma_{12}}
\end{align}
is nonzero for coinciding rapidities, namely $\tilde{h}(u,u)\ne 0$. The matrix parts of the two hexagons are in general complicated functions of $S$-matrix elements. The Zhukowsky variables are invariant under $\pm 4\gamma$ transformation, hence we have $\rH^{\text{mat}}(\{u\}_n|\{v^{-4\gamma}\}_n)=\rH^{\text{mat}}(\{u\}_n|\{v\}_n)$ and $\rH^{\text{mat}}(\{v^{4\gamma}\}_n|\{u\}_n)=\rH^{\text{mat}}(\{v\}_n|\{u\}_n)$. Finally we notice that when taking diagonal limit $v_i\to u_i$, due to the structure of the dynamical parts, poles appear. But since the structure constant is a well-defined object in diagonal limit, these poles necessarily cancel.

\section{The diagonal form factor}
In this section, we take the diagonal limit of the structure constant as discussed in the previous section. Let us denote the structure constant in the sum-over-partition formula (\ref{eq:sop}) as $\mathcal{C}_{2N}(\{u\}_N|\{v\}_N)$. Taking the diagonal limit we can define
\begin{align}
\label{eq:CN}
\mathcal{C}_{\mathrm{HHL}}(\{u\}_N)=\lim_{\epsilon_i\to 0}\mathcal{C}_{2N}(\{u_i\}_N|\{u_i+\epsilon_i\}_N).
\end{align}
We also define a quantity
\begin{align}
\label{eq:FN}
\mathcal{F}(\{u\}_N)=\lim_{\epsilon_i\to 0}\mathcal{C}'_{2N}(\{u_i\}_N|\{u_i+\epsilon_i\}_N),
\end{align}
where by prime we mean that before taking the limit, we replace all the factors $e^{i p(v_i) l}$ by $e^{i p(u_i) l}$.
In both cases, after taking the diagonal limit, we apply BAE in order to eliminate the $l$-dependence due to the phase factors $e^{i p(u_i) l}$.

The main result of our paper is the proof of the following statement:

\begin{theorem}
The heavy-heavy-light symmetric structure constant has the following $L$-dependence
\begin{align}
\label{theor}
\mathcal{C}_\mathrm{HHL}(\{u_i\}_N)=\frac{1}{\prod_{i=1}^N\tilde{h}(u_i,u_i)} \sum_{\alpha\cup\bar{\alpha}=\{u\}_N}F^s(\alpha)\rho^s_{L}(\bar{\alpha}),
\end{align}
where $L$ is the length of the heavy operator.
\end{theorem}
Here $\rho^s_L$ is the Jacobian in the symmetric scheme \cite{Pozsgay:2007gx} which is defined as
\begin{align}
&\,\Phi_j=p_jL-i\sum_{k\ne j}\log S(u_j,u_k),\\\nonumber
&\,\rho_L^s(\{u\}_N)=\det_{j,k}\frac{\partial\Phi_k}{\partial u_j}.
\end{align}
For subsets $\bar{\alpha}\subset\{u\}_N$, $\rho^s_L(\bar\alpha)$ is defined with respect to the rapidities $u_j\in\bar{\alpha}$. The $S$-matrix $S(u,v)$ is the all-loop $S$-matrix in the $\mathfrak{su}(2)$ sector, and
\begin{align}
F^s(\alpha)=\sum_{\beta\cup\bar{\beta}=\alpha}\mathcal{F}(\beta)\rho_{-l_0}^s(\bar{\beta}).
\end{align}

\subsection{The factorization property}
In order to prove the theorem, it is important to know the behavior of the hexagon form factor when two rapidities on the two edges coincide with each other. In this case, the resulting hexagon form factor is proportional to the hexagon form factors with less excitations. We will refer to this kind of relations as the factorization properties. In our case it can be summarized as the follows
\begin{align}
\label{eq:fac}
&\rH^{\text{mat}}(\textcolor{red}{u},\{u\}_n|\{v\}_n,\textcolor{red}{u})=-\rH^{\text{mat}}(\{u\}_n|\{v\}_n),\\\nonumber
&\rH^{\text{mat}}(\star,u,v,\star|\star)=S(u,v)\frac{h(v,u)}{h(u,v)}\rH^{\text{mat}}(\star,v,u,\star|\star),\\\nonumber
&\rH^{\text{mat}}(\star|\star,u,v,\star)=S(u,v)\frac{h(v,u)}{h(u,v)}\rH^{\text{mat}}(\star|\star,v,u,\star).
\end{align}
Note that here the relations only concern the matrix part. The first relation is a reformulation of the decoupling condition in \cite{Basso:2015zoa} where the authors performed a $2\gamma$ transformation. The scattering of a particle and its anti-particle gives rise to the singlet state \cite{Beisert:2005tm} which scatters trivially with any other excitations and thus can be factorized. This statement is equivalent to (\ref{eq:fac}) in the $-4\gamma$ transformation. The remaining two equations can be derived simply from the prescription for computing the matrix part and using the fact that the $\mathfrak{su}(2)$ $S$-matrix at all loop is given by
\begin{align}
S(u,v)=A(u,v)^2\frac{h(u,v)}{h(v,u)}
\end{align}
where $A(u,v)$ is one of Beisert's $S$-matrix element \cite{Beisert:2005tm}.

\subsection{A recursion relation}
As we mentioned above, the explicit $L$-dependence of the structure constant appears due to taking derivatives of the phase factor $e^{i l p(v_i)}$. This implies that the polynomial $l$ dependence always come within the quantity $z_i=lp'(u_i)$. Instead of studying directly the dependence on $L$, it is more convenient to consider the dependence on $z_i$'s.

Using the factorization property, we first prove the following lemma.
\begin{lemma}[Recursion relation]
The dependence of $\mathcal{C}_{\mathrm{HHL}}$ on $z_k$ is linear and satisfies the following relation
\begin{align}
\label{lem}
\frac{\partial}{\partial z_k}\mathcal{C}_{\mathrm{HHL}}(\{u\}_N)=a_k\,\mathcal{C}^{\mathrm{mod}}_{\mathrm{HHL}}(\{u\}_N/u_k)
\end{align}
where by the set $\{u\}_N/u_k$ we mean that the rapidity $u_k$ is excluded from the initial set and
\begin{align}
\label{eq:aN}
a_k=\frac{1}{\tilde{h}(u_k,u_k)}
\end{align}
and the index ``mod" stands for the following replacement
\begin{align}
\label{mod}
z_i \to z_i^{\mathrm{mod}}=z_i+\varphi(u_i,u_k).
\end{align}
$\varphi(u,v)$ is defined as $\varphi(u,v)=-i\frac{\partial}{\partial u}\log S(u,v)$.
\end{lemma}
\textbf{Proof} Let's first consider $\mathcal{C}_{2N}(\{u\}_N|\{v\}_N)$ and its dependence on $z_N$ after taking the limit $v_N\to u_N$. Among the terms in the sum-over-partition formula, the terms leading to the $z_N$ dependence have the factor
\begin{align}
\frac{e^{-ilp(u_N)+ilp(v_N)}}{h(u_N,v_N)}\times\cdots.
\end{align}
These are the terms with $u_N\in\bar{\alpha}$ and $v_N\in\bar{\beta}$ (both of the corresponding excitations are located on the second hexagon), which take the following form (begining from here and till the end of the subsection B we redefine the sets $\bar\alpha$, $\bar\beta$ as follows $\bar\alpha \to u_N \cup \bar\alpha$ and $\bar\beta \to \bar\beta \cup v_N$)
\begin{align}
&t(\alpha,\beta,u_N \cup \bar{\alpha},\bar{\beta}\cup v_N)=\\\nonumber
&\,\omega_{-l}(\alpha,u_N \cup \bar{\alpha})\omega_l(\beta,\bar{\beta}\cup v_N)
\,\times \rH(\alpha|\beta)\rH(\bar{\beta},v_N|u_N,\bar{\alpha}).
\end{align}
For the splitting factor, we have
\begin{align}
\label{eq:omega}
\frac{\omega_{-l}(\alpha,u_N \cup \bar{\alpha})\omega_{l}(\beta,\bar{\beta}\cup v_N)}{\omega_{-l}(\alpha,\bar{\alpha})\omega_{l}(\beta,\bar{\beta})}
=e^{-ilp(u_N)+ilp(v_N)}.
\end{align}
For the dynamical part of the hexagon form factor, we have
\begin{align}
\label{eq:Hd}
\frac{\rH^{\text{dyn}}(\bar{\beta},u_N|u_N,\bar{\alpha})}{\rH^{\text{dyn}}(\bar{\beta}|\bar{\alpha})}=
\frac{h(\bar{\beta},u_N)}{h(u_N,\bar{\beta})}
\frac{h(u_N,\bar{\alpha})}{h(\bar{\alpha},u_N)}\frac{1}{h(u_N,v_N)}
\end{align}
For the matrix part, we can use the factorization property (\ref{eq:fac})
\begin{align}
\label{eq:Hm}
\frac{\rH^{\text{mat}}(\bar{\beta},u_N|u_N,\bar{\alpha})}{\rH^{\text{mat}}(\bar{\beta}|\bar{\alpha})}=-A(\bar{\beta},u_N)^2A(u_N,\bar{\alpha})^2.
\end{align}
Combining the results (\ref{eq:omega}), (\ref{eq:Hd}) and (\ref{eq:Hm}), one can derive that
\begin{align}
&\frac{\partial}{\partial z_N}\left.t(\alpha,\beta,u_N \cup\bar{\alpha},\bar{\beta}\cup u_N+\epsilon_N)\right|_{\epsilon_N\to 0}\\\nonumber
&=a_N\frac{S(\bar{\beta},u_N)}{S(\bar{\alpha},u_N)}t(\alpha,\beta,\bar{\alpha},\bar{\beta})
\end{align}
where $a_N$ is given in (\ref{eq:aN}) and does not depend on the partition. We can combine $S(\bar{\alpha},u_N)$ and $S(\bar{\beta},u_N)$ with the splitting factors $\omega_{-l}(\alpha,\bar{\alpha})$ and $\omega_l(\beta,\bar{\beta})$ respectively. This leads to a modification of the momenta
\begin{align}
\label{repl}
&e^{-ilp(\bar{\alpha}_k)}\to e^{-ilp^{\text{mod}}(\bar{\alpha}_k)}=e^{-ilp(\bar{\alpha}_k)}S(u_N,\bar{\alpha}_k),\\\nonumber
&e^{ilp(\bar{\beta}_k)}\to e^{ilp^{\text{mod}}(\bar{\beta}_k)}=e^{ilp(\bar{\beta}_k)}S(\bar\beta_k,u_N).
\end{align}
with the form of the splitting factor (\ref{eq:sfactor}) unchanged. Therefore we can write
\begin{align}
&\frac{\partial}{\partial z_N}\left.t(\alpha,\beta,u_N \cup\bar{\alpha},\bar{\beta}\cup u_N+\epsilon_N)\right|_{\epsilon_N\to 0}\\\nonumber
&=a_N\,t(\alpha,\beta,\bar{\alpha},\bar{\beta})^{\text{mod}},
\end{align}
where again the index ``mod" implies the replacement (\ref{repl}).
Summing over partitions we get
\beq
\begin{split}
&\frac{\partial}{\partial z_N}\lim_{\epsilon_N \to 0}\mathcal{C}_{2N}(\{u\}_N|\{v\}_N)|_{v_N=u_N+\epsilon_N} =\\ &a_N\,\mathcal{C}^{\text{mod}}_{2N-2}(\{u\}_N/u_N|\{v\}_N/v_N).
\end{split}
\eeq
After taking the limit $v_i \to u_i$ for the rest of the rapidities we get
\begin{align}
\frac{\partial}{\partial z_N}\mathcal{C}_{\mathrm{HHL}}(\{u\}_N)=a_N\,\mathcal{C}^{\mathrm{mod}}_{\mathrm{HHL}}(\{u\}/u_N).
\end{align}
Finally, since the structure constant is symmetric with respect to the rapidities, we proved the lemma \ur{lem}.

\subsection{Proof of the theorem}
Now we are ready to prove the main theorem \ur{theor}. For a given partition $\alpha\cup\bar{\alpha}=\{u\}_N$, let us define
\begin{align}
\label{eq:KK}
K_N=\frac{1}{\prod_{k=1}^N\tilde{h}(u_k,u_k)}\,\rho^s_{l}(\bar{\alpha}).
\end{align}
First we prove that
\beq
\begin{split}
\label{state}
&\mathcal{C}_{\mathrm{HHL}}(\{u\}_N)=\\
&\sum_{\alpha\cup\bar{\alpha}=\{u\}_N}\mathcal{F}(\alpha)\,K_N(\bar{\alpha})
\equiv \mathcal{W}_N(\{u\}_N).
\end{split}
\eeq
Noticing that
\begin{align}
\frac{\partial}{\partial z_k}\rho_l^s(\{u\}_N)=\rho_{l}^{s,\text{mod}}(\{u\}_{N-1}),
\end{align}
with the modification rule \ur{mod}, we have
\beq
\label{Wlem}
\frac{\partial}{\partial z_k}\mathcal{W}_N(\{u\}_N)=a_k\,\mathcal{W}^{\text{mod}}_{N-1}(\{u\}_{N}/u_k).
\eeq
We prove (\ref{state}) by induction. The case $n=1$ can be verified easily by explicit computation. Assume that the \ur{state} is true for $n\le N-1$. From \ur{lem} and \ur{Wlem}, we find that the $z_i$ dependencies of $\mathcal{C}_{\mathrm{HHL}}(\{u\}_N)$ and $\mathcal{W}_N(\{u\}_N)$ are the same. In order to prove \ur{state} we simply need to show that the terms which are independent of $z$'s are equal.
Putting $z_i\to 0$ in \ur{state}, all the $\rho^s_l$ vanish and we have
\beq
\mathcal{W}_{N}(\{u\}_N)|_{z_i\to 0} = \mathcal{F}(\{u\}_N).
\eeq
On the other hand, from the definition of $\mathcal{F}_N(\{u\}_N)$, we first put $e^{i l p(v_i)}$ to $e^{i l p(u_i)}$ and then take the diagonal limit. This prevents the appearance of $z_i=l p'(u_i)$ dependent terms. Thus we have shown that
\beq
\mathcal{C}_{\text{HHL}}(\{u\}_N)=\frac{1}{\prod_{i=1}^N\tilde{h}(u_i,u_i)} \sum_{\alpha\cup\bar{\alpha}=\{u\}_N}\mathcal{F}(\alpha)\rho^s_{l}(\bar{\alpha}).
\eeq
Finally we go from $\rho^s_{l}$ to $\rho^s_{L}$, which can be done by the following relation:
\begin{align}
\rho^s_{l_1+l_2}(\{u\}_N)=\sum_{\alpha\cup\bar{\alpha}=\{u\}_N}\rho^s_{l_1}(\alpha)\rho^s_{l_2}(\bar{\alpha}).
\end{align}
Taking $l_1=L$ and $l_2=-l_0$, we have
\begin{align}
\mathcal{C}_{\text{HHL}}(\{u\}_N)=\frac{1}{\prod_{i=1}^N\tilde{h}(u_i,u_i)}\sum_{\alpha\cup\bar{\alpha}=\{u\}_N}F^s(\alpha)\,\rho^s_{L}(\bar{\alpha}),
\end{align}
where
\begin{align}
F^s(\alpha)=\sum_{\beta\cup\bar{\beta}=\alpha}\mathcal{F}(\beta)\rho_{-l_0}^s(\bar{\beta}).
\end{align}
This proves the theorem.

The expansion \ur{theor} can also be written in the so-called connected scheme (see \cite{Pozsgay:2007gx})
\begin{align}
\mathcal{C}_{\text{HHL}}(\{u\}_N)=\frac{1}{\prod_{i=1}^N\tilde{h}(u_i,u_i)}\sum_{\alpha\cup\bar{\alpha}=\{u\}_N}F^c(\alpha)\,\rho^c_{L}(\bar{\alpha})
\end{align}
where $\rho^c_L(\alpha)$ is defined to be the diagonal minor of the Jacobian $\rho^c_L(\{u\}_N)$ and hence depends on all the rapidities. The relations between $F^c$ and $F^s$ can be worked out explicitly \cite{Pozsgay:2007gx}.

Finally it is worth to mention that $\tilde{h}(u,u)=\frac{1}{\mu_X(u)}$, where $\mu_X(u)$ is the measure introduced in \cite{Basso:2015zoa}. It means that by normalizing the structure constant with the norm of the heavy operator we can get rid of these factors. The normalized structure constant is given by
\begin{align}
C_{\text{HHL}}(\{u\}_N)=\frac{1}{\rho^s_{L}(\{u\}_N)}\sum\limits_{\alpha\cup\bar{\alpha}=\{u\}_N}F^s(\alpha)\,\rho^s_{L}(\bar{\alpha}).
\end{align}
\section{Coefficients $F^s$}
According to the original proposal of \cite{Pozsgay:2007gx} the coefficients in the finite volume expansion are identified with the infinite volume form factors. We find it might be important to keep this identification in mind in the case of ${\cal N}=4$ SYM as well.

Using the hexagon approach, we compute these coefficients at all loops. We expand the results at weak coupling and compare with the ones computed in \cite{Hollo:2015cda} at tree level. We compare the results in the connected scheme for the case $l_0=1$ for a few magnons and obtain a perfect match. At tree level, these coefficients of $N$ excitations are conjectured to take the following form

\begin{align}
F^{c(0)}(\{u\}_N)=&\,\sigma^{(0)}_1\varphi^{(0)}_{12}\varphi^{(0)}_{23}\cdots\varphi^{(0)}_{N-1,N}
+\text{permutations}.
\end{align}
where
\begin{align}
\sigma^{(0)}(u)=\frac{1}{u^2+1/4},\quad \varphi^{(0)}(u,v)=\frac{2}{(u-v)^2+1}.
\end{align}
Interestingly, one loop computation indicates that the form still holds with the following corrections
\begin{align}
&\,\sigma^{(1)}(u)=\frac{1}{u^2+1/4}+\frac{8g^2\,u^2}{(u^2+1/4)^3},\\\nonumber
&\,\varphi^{(1)}(u,v)=\frac{2}{(u-v)^2+1}\\\nonumber
&\,\qquad+\frac{4g^2(u^2-v^2)}{(u^2+1/4)(v^2+1/4)((u-v)^2+1)},
\end{align}
which we checked up to three magnons.
It is possible that the same ansatz will hold at even higher orders with proper modifications. If this is the case, it will simplify a lot the computation of these coefficients. We leave this interesting problem for the future investigation. It would also be interesting to understand better the physical meaning of these coefficients in the context of ${\cal N}=4$ SYM.

\section{Conclusion and discussions}
We study the symmetric heavy-heavy-light structure constant within the hexagon approach and proved that its asymptotic $L$-dependence, where $L$ is the length of the heavy operators, is given by the expansion \ur{theor}. This study is for the special case where the excitations of the heavy operators are the transverse scalar excitations $X$,$\bar{X}$ and the light operator being the rotated BMN vacuum. However, the methods of the present paper can be generalized straightforwardly to the cases where the excitations on the heavy operator are transverse derivatives or the longitudinal excitations and the light operator being non-BPS operator. Finally, in order to have a complete proof of the BJW conjecture within the hexagon approach, we also need to consider the bridge wrapping. All these issues will be treated in more detail in the forthcoming paper \cite{HJP}.

\section*{Acknowledgement}
It's our pleasure to thank Zoltan Bajnok, Ivan Kostov, Didina Serban and especially Benjamin Basso and Shota Komatsu for many valuable discussions, correspondences and comments on the manuscript. We also thank Laszlo Hollo for initial collaboration on the project. The work of A.P. receives support from the People Programme (Marie Curie Actions) of the European Union's Seventh
Framework Programme FP7/2007-2013/ under REA Grant Agreement No 317089.

\providecommand{\href}[2]{#2}\begingroup\raggedright\endgroup

\bibliographystyle{utphys}
\bibliography{yunfeng}

\begin{thebibliography}{10}

\bibitem{Basso:2015zoa}
B.~Basso, S.~Komatsu, and P.~Vieira, ``{Structure Constants and Integrable
  Bootstrap in Planar N=4 SYM Theory},''
\href{http://arxiv.org/abs/1505.06745}{{\ttfamily arXiv:1505.06745 [hep-th]}}.

\bibitem{Bajnok:2014sza}
Z.~Bajnok, R.~A. Janik, and A.~Wereszczyski, ``{HHL correlators, orbit
  averaging and form factors},''
  \href{http://dx.doi.org/10.1007/JHEP09(2014)050}{{\em JHEP} {\bfseries 1409}
  (2014) 050},
\href{http://arxiv.org/abs/1404.4556}{{\ttfamily arXiv:1404.4556 [hep-th]}}.

\bibitem{Zarembo:2010ab}
K.~{Zarembo}, ``{Holographic three-point functions of semiclassical states},''
  \href{http://dx.doi.org/10.1007/JHEP09(2010)030}{{\em Journal of High Energy
  Physics} {\bfseries 9} (Sept., 2010) 30},
  \href{http://arxiv.org/abs/1008.1059}{{\ttfamily arXiv:1008.1059 [hep-th]}}.

\bibitem{Costa:3pt}
M.~S. {Costa}, R.~{Monteiro}, J.~E. {Santos}, and D.~{Zoakos}, ``{On
  three-point correlation functions in the gauge/gravity duality},''
  \href{http://dx.doi.org/10.1007/JHEP11(2010)141}{{\em JHEP} {\bfseries 11}
  (2010) 141}, \href{http://arxiv.org/abs/1008.1070}{{\ttfamily
  arXiv:1008.1070}}.

\bibitem{Pozsgay:2007gx}
B.~Pozsgay and G.~Takacs, ``{Form factors in finite volume. II. Disconnected
  terms and finite temperature correlators},''
  \href{http://dx.doi.org/10.1016/j.nuclphysb.2007.07.008}{{\em Nucl.Phys.}
  {\bfseries B788} (2008) 209--251},
\href{http://arxiv.org/abs/0706.3605}{{\ttfamily arXiv:0706.3605 [hep-th]}}.

\bibitem{Hollo:2015cda}
L.~Hollo, Y.~Jiang, and A.~Petrovskii, ``{Diagonal Form Factors and
  Heavy-Heavy-Light Three-Point Functions at Weak Coupling},''
\href{http://arxiv.org/abs/1504.07133}{{\ttfamily arXiv:1504.07133 [hep-th]}}.

\bibitem{HJP}
Y.~Jiang, ``{Diagonal Form Factors and Hexagon Form Factors II. Non-BPS Light
  Operator},''
\href{http://arxiv.org/abs/1601.06926}{{\ttfamily arXiv:1601.06926 [hep-th]}}.

\bibitem{Beisert:2006ez}
N.~Beisert, B.~Eden, and M.~Staudacher, ``{Transcendentality and Crossing},''
  \href{http://dx.doi.org/10.1088/1742-5468/2007/01/P01021}{{\em J.Stat.Mech.}
  {\bfseries 0701} (2007) P01021},
\href{http://arxiv.org/abs/hep-th/0610251}{{\ttfamily arXiv:hep-th/0610251
  [hep-th]}}.

\bibitem{Beisert:2005tm}
N.~Beisert, ``{The $\mathfrak{su}(2|2)$ dynamic S-matrix},''
  \href{http://dx.doi.org/10.4310/ATMP.2008.v12.n5.a1}{{\em
  Adv.Theor.Math.Phys.} {\bfseries 12} (2008) 945--979},
\href{http://arxiv.org/abs/hep-th/0511082}{{\ttfamily arXiv:hep-th/0511082
  [hep-th]}}.

\bibitem{Beisert:2006qh}
N.~Beisert, ``{The Analytic Bethe Ansatz for a Chain with Centrally Extended
  su(2|2) Symmetry},''
  \href{http://dx.doi.org/10.1088/1742-5468/2007/01/P01017}{{\em J. Stat.
  Mech.} {\bfseries 0701} (2007) P01017},
\href{http://arxiv.org/abs/nlin/0610017}{{\ttfamily arXiv:nlin/0610017
  [nlin.SI]}}.

\end{thebibliography}

\end{document}